\documentclass{ipart_v1}

\usepackage{t1enc}
\usepackage[utf8]{inputenc}
\usepackage[english]{babel}

\usepackage{amsthm}
\usepackage{yfonts}

\usepackage{bbm}
\usepackage{bm}
\usepackage{mathrsfs}
\usepackage{comment}

\newcommand{\be}[0]{\begin{equation}}
\newcommand{\ee}[0]{\end{equation}}

\numberwithin{equation}{section}

\theoremstyle{plain}

\usepackage{bm}
\begin{document}

\author{Dalibor Jav\r{u}rek}
\email{dalibor.javurek@mojeposta.xyz}

\title[Energy Momentum Tensor of WGSQM Theory]{Energy Momentum Tensor of Extended and Non-extended Theory of Weak Gravity and Spinor Quantum Mechanics}

\date{\today}

\begin{abstract}
Two distinct energy-momentum tensors of the theory of weak gravity and spinor quantum mechanics are analyzed with respect to their four-divergence and expectation values of energy. The first energy-momentum tensor is obtained by a straightforward generalization of the symmetric energy-momentum tensor of a free Dirac field, and the second is derived by the second Noether theorem. We find that the four-divergences of both tensors are not equal. Particularly, the tensor derived by the generalization procedure does not match the four-divergence of the canonical energy-momentum tensor. As a result, both tensors predict distinct values for the energy of the Dirac field. The energy-momentum tensor of the non-extended theory with the correct expression for four-divergence obtained by the second Noether theorem is asymmetric. This contradicts the requirements of general relativity. To rectify this situation, the Lagrangian of the theory is extended with the Lagrangian of the free electromagnetic field on curved spacetime. Then, the symmetric energy-momentum tensor of quantum electrodynamics with the required four-divergence is obtained by the second Noether theorem. Moreover, the energy-momentum tensor appears in the interaction Lagrangian term of the extended theory. In addition, we show that the Lagrangian density of the extended theory can be recast into the Lagrangian density of a flat spacetime theory, contrary to the statement made for the non-extended theory.
\end{abstract}


\maketitle

\section{Introduction}

In the realm of quantum systems interacting with gravitational fields, notable experiments and theoretical studies have been conducted in relation to Earth's gravitational influence. These investigations include studies of neutrons interacting with the Earth's Newtonian field~\cite{Nesvizhevsky2002,Tobar2022} as well as atom interferometer experiments aimed at rigorously testing the equivalence principle~\cite{Zych2015}. Moreover, a gravitational wave detector composed of two separated atom interferometers has been proposed~\cite{Dimopoulos2008}. Although attempts to discern gravitomagnetic effects with these devices have been discussed, the formidable challenges arising from the effects' small scale make such experiments inherently difficult~\cite{Adler2012}.

Recent research on Dirac fields in curved space-time has expanded our knowledge in several ways. A rigorous, yet still pedagogically valuable study has looked into how quantized Dirac fields behave in curved space-times with low relativistic effects \cite{Falcone2022}. The influence of two-dimensional space curvature on massless Dirac particles has been examined \cite{Flouris2018}, and the criteria for quantizing Dirac fields in curved space-times have been set \cite{Cortez2020}. Attention has been devoted to the exploration of quantum walks related to massless Dirac fermions in curved space-time \cite{Dimolfetta2013} and to entropic dynamics \cite{Ipek2019}. Detailed analyses of solutions and oscillator models in curved environments have been performed \cite{Oliveira2019, Oliveira2020,Yagdjian2021}. The research demonstrates the ongoing effort to understand how quantum fields interact with the structure of space-time.

Many other interesting effects resulting from the gravitational field coupling to the Dirac field have been studied. Attention has been devoted to the investigation of parity violation due to the Dirac field's gravity-spin interaction~\cite{Jentschura2013}. The interplay between Einstein's equivalence principle, atomic transition frequencies, and the g-factor measurements has been explored~\cite{Jentschura2018}. The symmetry guaranteeing the same gravitational attraction for electrons and positrons has been found~\cite{Jentschura2013b}. Despite all the knowledge already obtained about the topic, we believe there are still remarkable questions and areas yet unexplored.

The energy-momentum tensor is an important object in quantum theory~\cite{Kosyakov2007,Schwartz2013}. It allows for the calculation of expectation values of mechanical quantities like energy, momentum, pressure, or angular momentum. Moreover, in complex systems, it provides insight into the distribution of energy and momentum among individual parts of the system. In the general theory of relativity, the energy-momentum tensor plays a crucial role since it serves as a source of spacetime curvature~\cite{Carroll2019,Misner1973}. The energy-momentum tensor in quantum field theory on flat spacetime is usually calculated by the first or second Noether theorem~\cite{Kosyakov2007,Freese2022}. In general relativity, the Hilbert energy-momentum tensor for a physical system involving spacetime curvature is calculated by variation of the Lagrangian density with respect to metric coefficients~\cite{Carroll2019}. In the GQM theory, which is formulated for a weak gravitational field, the flat spacetime energy-momentum tensor is a sufficient approximation of the exact energy-momentum tensor~\cite{Adler2012}.

In this manuscript, we identify that the symmetric energy-momentum tensor of the GQM theory~\cite{Adler2012}, obtained by a generalization of the free Dirac field energy-momentum tensor, is not equal to the energy-momentum tensor obtained by the second Noether theorem. We show that the four-divergence of the generalized energy-momentum tensor does not match the four-divergence of the canonical energy-momentum tensor. As a result, the values of the four-momentum predicted by both tensors are not the same. This is shown explicitly for the energy. The energy-momentum tensor with the correct four-divergence is asymmetric. In consequence, the symmetric tensor can no longer be considered as valid energy-momentum tensor.

The symmetric tensor appears in Lagrangian of the GQM theory in a role of source of the gravitational field. Therefore, it is necessary to alter the theory to contain a valid energy-momentum tensor in the source term. To obtain a symmetric energy-momentum tensor with the correct value of four-divergence, it is necessary to extend the Lagrangian density of the GQM theory by the Lagrangian density of the free electromagnetic field. After the extension the theory can be interpreted as a flat spacetime theory.

After an introduction to the topic is provided in Sec.~1, we summarize the GQM theory together with its energy-momentum tensor in Sec.~2. In Sec.~3, we compare this energy-momentum tensor with the energy-momentum tensor derived by the second Noether theorem. In addition, we compare the four-divergences of both tensors and analyze the expressions for the energy of both tensors. In Sec.~4, the Lagrangian of the GQM theory is extended with a term related to the free electromagnetic field. It is shown that the addition of this term is necessary to obtain a symmetric energy-momentum tensor in the interaction Lagrangian density. In Sec.~5, we show that the Lagrangian density of the extended GQM theory can be recast as the Lagrangian density of a flat spacetime theory. Then, the gravitational field is treated as a tensor field with a source in form of energy-momentum tensor of quantum electrodynamics. Sec.~6 summarizes all findings mentioned in the paper.

\section{Theory of Gravitomagnetism and Spinor Quantum Mechanics}
A theory describing the interaction between the electromagnetic four-potential field $A^\mu$ and the Dirac spinor field $\psi$ on a weakly curved spacetime has been developed by Adler \textit{et al.} \cite{Adler2012}. Since the gravitational field is assumed to be weak, the coefficients of the metric tensor $g_{\mu \nu}$ are defined by the equation
\begin{equation}
\label{e01}
g_{\mu \nu} = \eta_{\mu \nu} + h_{\mu \nu}.
\end{equation}
$\eta_{\mu \nu} = \mathrm{diag}(1,-1,-1,-1)$ is the metric tensor of flat spacetime, and the coefficients of the tensor $h_{\mu \nu}$ describe small metric perturbations on an otherwise flat spacetime. The coefficients $h_{\mu \nu}$ are assumed to be much smaller than one, $ \vert h_{\mu \nu} \vert \ll 1$ in absolute value. Geometrized units $c = \hbar = \mu_0 = \varepsilon_0 = 1$ are used in the paper.

The theory~\cite{Adler2012} has been developed with the principle of general covariance. Specifically, the GQM theory is based upon the generalization of the flat space-time Lagrangian density of the Dirac field $\psi$ minimally coupled to the electromagnetic field $A^\mu$ to a general spacetime. The GQM theory assumes, that the gravitational field is weak and the sources of the gravitational field are moving slowly. After simplification and rearrangement of terms in the Lagrangian of the theory $L$, it has been shown that the Lagrangian $L$ attains the form
\begin{align}
\label{e1}
 L &= L_{\rm flat} + L_{\rm IG}\\
 \label{e2}
 L_{\rm flat} &=  \dfrac{i}{2} [\bar{\psi} \gamma^\mu (\partial_\mu \psi) - (\partial_\mu \bar{\psi}) \gamma^\mu \psi] - m \bar{\psi} \psi - q A_\mu \bar{\psi} \gamma^\mu \psi \\
\nonumber
L_{\rm IG} &= -\dfrac{h_{\mu \nu}}{4} \left[ \bar{\psi} \gamma^\nu(i \partial^\mu \psi - q A^\mu \psi) - (i \partial^\mu \bar{\psi} + qA^\mu \bar{\psi}) \gamma^\nu \psi \right] \\
\label{e3}
&= -\dfrac{1}{2} h_{\mu \nu} T^{\mu \nu};
\end{align}
\begin{align}
\nonumber
T^{\mu \nu}&= \dfrac{i}{4}(\bar{\psi} \gamma^\mu \partial^\nu \psi - \partial^\nu \bar{\psi} \gamma^\mu \psi + \bar{\psi} \gamma^\nu \partial^\mu \psi - \partial^\mu \bar{\psi} \gamma^\nu \psi) + \\
\label{e31}
&-\dfrac{q}{2}(\bar{\psi} A^\nu \gamma^\mu \psi + \bar{\psi} A^\mu \gamma^\nu \psi).
\end{align}
The $L_{\rm flat}$ is the Lagrangian of the Dirac field $\psi$ minimally coupled to the electromagnetic field $A^\mu$ on flat spacetime; $L_{\rm IG}$ is the interaction Lagrangian describing the interaction of the gravitational field $h_{\mu \nu}$ with the electromagnetic field $A^\mu$ and the Dirac field $\psi$; $\gamma^\mu$ are flat spacetime Dirac matrices obeying the anti-commutation relation
\begin{equation}
[\gamma^\mu,\gamma^\mu]_+ = \gamma^\mu \gamma^\nu + \gamma^\nu \gamma^\mu= 2 \eta^{\mu \nu}.
\end{equation}
The field $\bar{\psi}$ is related to the Dirac field $\psi$ through the relation $\bar{\psi} = \psi^\dagger \gamma^0$. It has been shown~\cite{Adler2012} that the interaction Lagrangian $L_{\rm IG}$ can be written as a scalar product between the gravitational field tensor $h_{\mu \nu}$ and the tensor $T^{\mu \nu}$, in the same manner as the interaction Lagrangian between the electromagnetic field $A^\mu$ and the electric four-current $J^\mu = q\bar{\psi} \gamma^\mu \psi$. The tensor $T^{\mu \nu}$ has been identified with the energy-momentum tensor of the Dirac field minimally coupled to the electromagnetic field $A^\mu$ on flat spacetime~\cite{Adler2012}. It has been derived by a generalization of the energy-momentum tensor for the free Dirac field on flat spacetime to the case where the electromagnetic field $A^\mu$ is present.

The action of the theory $S$ is given by a volume integral of the Lagrangian of the theory $L$ in Eq.~(\ref{e1}) over the whole spacetime, which is considered to be weakly curved.
\begin{equation}
\label{e32}
 S = \int d^4x \sqrt{-g}\, L.
\end{equation}
In Eq.~(\ref{e32}) $g$ denotes the determinant of the metric tensor $g_{\mu \nu}$. In accordance with Adler~et.~al., let us denote
the overall Lagrangian density under the integral in Eq.~(\ref{e32}) as
\begin{equation}
 \mathcal{L} = \sqrt{-g}\, L.
\end{equation}
Physically, the quantity $\mathcal{L}$ can be considered as the Lagrangian density of a theory on flat spacetime $x^\mu$. However, it must be kept in mind that the coordinates $x^\mu$ are related to physical distances through the metric tensor $g_{\mu \nu}$ in Eq.~(\ref{e01}).

At the end of this chapter, we mention a few important quantities that will be used in this work. The tensor of the electromagnetic field $F_{\mu \nu}$ on curved spacetime is defined with covariant derivatives $\nabla_\mu$ by the equation~\cite{Bunney2022}
\begin{equation}
\label{e13}
 F_{\mu \nu} = \nabla_\mu A_\nu - \nabla_\nu A_\mu = \partial_\mu A_\nu - \partial_\nu A_\mu.
\end{equation}
Due to the symmetry of the Christoffel symbols $\Gamma^{\nu}_{\mu \rho} = \Gamma^\nu_{\rho \mu}$~\cite{Bunney2022}, the electromagnetic field tensor $F_{\mu \nu}$ takes the same form as it does on flat spacetime. The generally covariant Lagrangian of the free electromagnetic field $L_{\rm EM}$ is given by the following equation
\begin{equation}
\label{e141}
 L_{\rm EM} = -\dfrac{1}{4} F^{\mu \nu} F_{\mu \nu} = -\dfrac{1}{4} g^{\mu \lambda} g^{\nu \rho} F_{\lambda \rho} F_{\mu \nu}.
\end{equation}
The energy-momentum tensor $T^{\mu \nu}_{\rm EM}$ of the free electromagnetic field on flat space-time is equal to
\begin{equation}
\label{e142}
 T^{\mu \nu}_{\rm EM} = F^{\mu}_{\phantom{\mu} \rho} F^{\rho \nu} + \dfrac{1}{4} \eta^{\mu \nu} F^{\rho \lambda} F_{\rho \lambda}.
\end{equation}

\section{Energy-Momentum Tensor of the Non-Extended Theory}

In this section, we prove that the energy-momentum tensor $T^{\mu \nu}$ of the GQM theory in Eq.~(\ref{e31}) cannot be related to the canonical energy-momentum tensor $\Theta^{\mu \nu}$ using the Belinfante procedure. We show that the energy-momentum tensor derived with the second Noether theorem, $T^{\mu \nu}_{\rm N}$, which satisfies this relation, is distinct and asymmetric. As a result, the tensors predict different energy values for the Dirac field $\psi$ in an external electromagnetic field $A^\mu$.

In article~\cite{Adler2012}, where the GQM theory has been developed, the authors derived the energy-momentum tensor $T^{\mu \nu}$ of the Dirac field minimally coupled to the electromagnetic field $A^\mu$, see appendix~B in~\cite{Adler2012}. They started with the derivation of the canonical energy-momentum tensor for the free Dirac field on flat space-time. Then, they correctly stated that the canonical energy-momentum tensor can be symmetrized into the form in Eq.~(B3) in \cite{Adler2012}.
To include the electromagnetic interaction into the energy-momentum tensor, they exchanged all the derivatives multiplied by an imaginary unit $i\partial_\mu$ in the tensor for $i\partial_\mu - q A_\mu$. They arrived at a symmetric energy-momentum tensor $T^{\mu \nu}$ in Eq.~(\ref{e31}), which can be rewritten into the form:
\begin{align}
\label{e4}
T^{\mu \nu} &= \dfrac{1}{2}\left( \Theta^{\mu \nu} + \Theta^{\nu \mu} -  q A^\mu \bar{\psi} \gamma^\nu \psi - q A^\nu \bar{\psi} \gamma^\mu \psi \right);\\
\label{e41}
\Theta^{\mu \nu} &= \dfrac{i}{2} \left[ \bar{\psi} \gamma^\mu (\partial^\nu \psi) -  (\partial^\nu \bar{\psi}) \gamma^\mu \psi \right],
\end{align}
where $\Theta^{\mu \nu}$ denotes the canonical energy-momentum tensor of the free Dirac field, which is not symmetric. The validity of the energy-momentum tensor in Eq.~(\ref{e4}) was justified by the calculation of its four-divergence
\begin{equation}
\label{e5}
\partial_\mu T^{\mu \nu} = -q(\bar{\psi} \gamma_{\mu} \psi) F^{\nu \mu} = - F^{\nu \mu} J_\mu.
\end{equation}
where $F^{\nu \mu}$ is the electromagnetic field tensor on flat space-time.

We derive the energy-momentum tensor $T^{\mu \nu}_{\rm N}$ from the Lagrangian $L_{\rm flat}$ in Eq.~(\ref{e2}) with the second Noether theorem~\cite{Freese2022}. The theorem does not have to provide a symmetric energy-momentum tensor for a physical system, as is the case here, but usually it does~\cite{Freese2022}. The energy-momentum tensor $T^{\mu \nu}_{\rm N}$ derived from the Lagrangian $L_{\rm flat}$ in Eq.~(\ref{e2}) with the second Noether theorem is given by the equation~\cite{Markoutsakis2019}
\begin{equation}
\label{e6}
T^{\mu \nu}_{\rm N} = \dfrac{1}{2}(\Theta^{\mu \nu} + \Theta^{\nu \mu} + q \bar{\psi} A^\nu \gamma^\mu \psi - q \bar{\psi} A^\mu \gamma^\nu \psi).
\end{equation}

The tensor $T^{\mu \nu}_{\rm N}$ in Eq.~(\ref{e6}) consists of a symmetric part $\Theta^{\mu \nu} + \Theta^{\nu \mu}$ and an anti-symmetric part, explicitly dependent on the components of the electromagnetic field $A^\mu$. Therefore, the obtained overall energy-momentum tensor $T^{\mu \nu}_{\rm N}$ is not symmetric and not equal to the energy-momentum tensor $T^{\mu \nu}$ obtained in Eq.~(\ref{e4}). Particularly, the derived energy-momentum tensor $T^{\mu \nu}_{\rm N}$ differs from the energy-momentum tensor $T^{\mu \nu}$ in Eq.~(\ref{e31}) in sign of its third term. Other terms of the energy-momentum tensors $T^{\mu \nu}$ and $T^{\mu \nu}_{\rm N}$ are identical. The asymmetry in the tensor $T^{\mu \nu}_{\rm N}$ occurs, because the electromagnetic field $A^\mu$ transfers its angular momentum to the Dirac field, but its intrinsic angular momentum is not accounted for in the tensor $T^{\mu \nu}_{\rm N}$. This issue is addressed in Section~5.

There are two distinct energy-momentum tensors calculated for the theory defined by the Lagrangian density $L_{\rm flat}$ in Eq.~(\ref{e2}). The symmetric one $T^{\mu \nu}$ in Eq.~(\ref{e4}), derived by the authors in~\cite{Adler2012}, and the asymmetric one $T^{\mu \nu}_{\rm N}$ in Eq.~(\ref{e6}), derived by the second Noether theorem. A crucial feature of the energy-momentum tensor $T^{\mu \nu}$ in Eq.~(\ref{e4}) is the inequality of its four-divergence, shown in Eq.~(\ref{e5}) compared to the four-divergence of the canonical energy-momentum tensor $\Theta^{\mu \nu}$~\cite{Kosyakov2007}:
\begin{equation}
\label{e61}
  \partial_\mu \Theta^{\mu \nu} = -\dfrac{\partial L}{\partial A_\mu} (\partial^\nu A_\mu) = q (\partial^\nu A_\mu) \bar{\psi} \gamma^\mu \psi.
\end{equation}
This occurs, because the Belinfante tensor $B^{\lambda \mu \nu}$ relating both tensors (by an equation analogous to Eq.~(\ref{e701})) cannot be found. Moreover, the tensors $\Theta^{\mu \nu}$ and $T^{\mu \nu}$ do not provide the same values of the four-momentum compoents $P^\mu$.

On the other hand, the four-divergence of the energy-momentum tensor $T^{\mu \nu}_{\rm N}$ is equal to the four-divergence of the canonical energy-momentum tensor $\Theta^{\mu \nu}$. This follows either from direct calculation or from the identification of the Belinfante tensor $B^{\lambda \mu \nu}$, which relates the energy-momentum tensor $T^{\mu \nu}_{\rm N}$ to the canonical energy-momentum tensor by the equation
\begin{align}
\label{e701}
T^{\mu \nu}_{\rm N} &= \Theta^{\mu \nu} + \partial_\lambda B^{\lambda \mu \nu};\\
\nonumber
B^{\lambda \mu \nu} &= \dfrac{i}{4} \left[ - \dfrac{\partial L}{\partial (\partial_\lambda \psi)} \sigma^{\mu \nu} \psi + \bar{\psi} \sigma^{\mu \nu} \dfrac{\partial L}{\partial(\partial_\lambda \bar{\psi})} \right]\\
\label{e702}
&= \dfrac{1}{8} \bar{\psi}[\gamma^\lambda,\sigma^{\mu \nu}]_+ \psi;\,\sigma^{\mu \nu} = \dfrac{i}{2}[\gamma^\mu,\gamma^\nu].
\end{align}
The anti-symmetry of the Belinfante tensor in its first two indices guarantees that the four-divergence of the energy-momentum tensor $T^{\mu \nu}_{\rm N}$ is equal to the four-divergence of the canonical energy-momentum tensor $\Theta^{\mu \nu}$.

Beacuse of the relation~(\ref{e701}), the energy-momentum tensor $T^{\mu \nu}_{\rm N}$ provides us the same values of four-momentum components $P^\mu$ as the canonical energy-momentum tesor $\Theta^{\mu \nu}$. This follows directly from definition of the four-momentum vector component $P^\mu$ by means of the energy-momentum tensor $T^{\mu \nu}_{\rm N}$
\begin{align}
\label{e703}
P^\mu &= \int d^3x\, T^{0 \mu}_{\rm N} = \\
\label{e704}
&= \int d^3 x\, \Theta^{0 \mu} + \int d^3 x\,\partial_0 B^{00 \mu} + \int d^3 x\,\partial_j B^{j 0 \nu}\\
\label{e705}
&= \int d^3 x\, \Theta^{0 \mu}.
\end{align}
The second term in Eq.~(\ref{e704}) vanishes because of the antisymmetry requirement on the Belinfante tensor $B^{\lambda \mu \nu} = -B^{\mu \lambda \nu}$ and the third term because of the utilization of the Gauss's law and the requirement that the resulting surface integral vanishes.

We show the distinctions of tensors $\Theta^{\mu \nu}$ and $T^{\mu \nu}$ with respect to provided components of the four-momentum vector $P^\mu$ on calculation of the energy $E$. The energy $E$ is derived with the energy-momentum tensor component $\Theta^{00}$ and $T^{00}$. Particularly, we focus on the calculation of the energy of the Dirac field $\psi$ in an external electromagnetic field $A^\mu$ and restrict ourselves to the case, when $A^0 = \varphi$ is not a function of time.  We will begin by calculating the energy of the system $E_\Theta$, using the component of the canonical energy-momentum tensor $\Theta^{00}$. The energy $E_{\Theta}$ is equal to
\begin{equation}
\label{e71}
E_{\Theta} = \int d^3 x\, \Theta^{00} = \int d^3 x\, \psi^\dagger (\hat{H} \psi),
\end{equation}
where $\hat{H}$ stands for the Hamiltonian operator of the Dirac field in an external electromagnetic field~\cite{Das2008}
\begin{equation}
 \hat{H} = \alpha^k(-i\partial_k - q A^k) + m \beta + q \varphi
\end{equation}
satisfying equation
\begin{equation}
\label{e72}
i \partial_0 \psi = \hat{H} \psi.
\end{equation}
A similar equation, but for $\psi^\dagger$, can be obtained by the Hermitian conjugation of the equation. As a result, the energy $E_{\Theta}$ is equal to the formula for the mean value of the energy of the Dirac field in the state $\psi$. We would arrive at the same expression for energy in Eq.~(\ref{e71}) if the $00$ component of the energy-momentum tensor $T^{\mu \nu}_{\rm N}$ was used. This is simply because $T^{00}_{\rm N} = \Theta^{00}$, as follows from Eq. (\ref{e6}).

On the other hand, if we calculate the value of the energy $E_T$ of the Dirac field from component $T^{00}$ we obtain
\begin{equation}
\label{e73}
 E_T = \int d^3 x\, T^{00} = \int d^3 x\, \Theta^{00} - q\int d^3 x\,\varphi \psi^\dagger \psi.
\end{equation}
The last term on the right-hand side of Eq.~(\ref{e73}) represents the negative value of the energy of the Dirac field in the scalar electric potential $\varphi$. As a result, the energy $E_\Theta$ in Eq.~(\ref{e71}) calculated by means of the $\Theta^{00}$ component is not equal to the energy $E_T$ calculated by means of the $T^{00}$ component. To be precise, it holds
\begin{equation}
\label{e76}
 E_T = E_\Theta -q\int d^3 x\,\varphi \psi^\dagger \psi.
\end{equation}
From a physical perspective, the second term on the right-hand side of Eq.~(\ref{e76}) subtracts the energy of the Dirac field in the scalar electric potential $\varphi$ from its total energy $E_\Theta$.

Overall, we argue, that the energy-momentum tensor $T^{\mu \nu}$ in Eq.~(\ref{e4}) is not a valid energy-momentum tensor of the GQM theory, because it is unable to provide the same value of energy as the canonical energy-momentum tensor $\Theta^{\mu \nu}$. The value of energy $E_\Theta$ is correct for two reasons. First, the canonical energy-momentum tensor is a Noether current associated with the global translational symmetries of the system. Second, it provides a physically meaningful formula for the energy of the Dirac field $\psi$ in the form of the mean value of the energy operator $\hat{H}$ (see Eq.~(\ref{e71})).

\section{Extension of the Theory with Lagrangian Density of Free Electromagnetic Field}

To obtain a symmetric energy-momentum tensor $T^{\mu \nu}_{\rm N}$  with the second Noether theorem from a theory based on the Lagrangian $L_{\rm flat}$ in Eq.~(\ref{e2}), it is necessary to extend the Lagrangian with the Lagrangian of the free electromagnetic field $L_{\rm EM}$ in Eq.~(\ref{e141}). This extension results in the Lagrangian of quantum electrodynamics on flat space-time~\cite{Freese2022}
\begin{equation}
\label{e8}
 L_{\rm QED, flat} = L_{\rm flat} + L_{\rm EM}.
\end{equation}
The energy-momentum tensor derived from the Lagrangian in Eq.~(\ref{e8}) with the second Noether theorem is equal to energy-momentum tensor of quantum electrodynamics~\cite{Freese2022}
\begin{align}
\nonumber
T^{\mu \nu}_{\rm QED} &= T^{\mu \nu}_{\rm N} + F^{\mu}_{\phantom{\mu} \lambda} F^{\lambda \nu} + \dfrac{\eta^{\mu \nu}}{4}F^{\rho \lambda} F_{\rho \lambda} - q A^\nu \bar{\psi}
\gamma^\mu \psi \\
\nonumber
&= \dfrac{1}{2}(\Theta^{\mu \nu} + \Theta^{\nu \mu}) - \dfrac{q}{2}( A^\nu \bar{\psi} \gamma^\mu \psi + A^\mu \bar{\psi} \gamma^\nu \psi) +  F^{\mu}_{\phantom{\mu} \lambda} F^{\lambda \nu}\\
\label{e9}
&- \eta^{\mu \nu} L_{\rm QED, flat}.
\end{align}

The tensor $T^{\mu \nu}_{\rm QED}$ is symmetric and its four-divergence is equal to the four-divergence of the canonical energy-momentum tensor of the theory
\begin{equation}
\Theta^{\mu \nu}_{\rm QED} = \Theta^{\mu \nu} -F^{\mu \lambda} \partial^\nu A_\lambda + \dfrac{1}{4}\eta^{\mu \nu} F^{\mu \nu} F_{\mu \nu}.
\end{equation}
The Belinfante tensor $B^{\lambda \mu \nu}_{\rm QED}$, which relates the symmetric energy-momentum tensor $T^{\mu \nu}_{\rm QED}$ with canonical energy-momentum tensor $\Theta^{\mu \nu}_{\rm QED}$ via relation analogical to~(\ref{e701}) equals
\begin{equation}
 B^{\lambda \mu \nu}_{\rm QED} = B^{\lambda \mu \nu} + F^{\mu \lambda} A^\nu.
\end{equation}
The four-divergence of the symmetric energy-momentum tensor $T^{\mu \nu}_{\rm QED}$ equals to zero.

According to the first line of Eq.~(\ref{e9}), the tensor can be decomposed into four parts. The first part, $T^{\mu \nu}_{N}$, represents the energy-momentum tensor derived with the second Noether theorem for the theory lacking the Lagrangian of the free electromagnetic field $L_{\rm EM}$, see Eq.~(\ref{e2}). The second and third terms in the first line of Eq.~(\ref{e9}) correspond to the energy-momentum tensor $T^{\mu \nu}_{\rm EM}$ of the free electromagnetic field, see Eq.~(\ref{e142}). The last term in the first line of Eq.~(\ref{e9}) covers the interaction energy of the electromagnetic field $A^\mu$ with the Dirac current $J^\mu~=~\bar{\psi} \gamma^\mu \psi$. The addition of the fourth term to the energy-momentum tensor $T^{\mu \nu}_{\rm N}$ creates a symmetric energy-momentum tensor $T^{\mu \nu}$ in Eq.~(\ref{e4})/\footnote{ The last term in Eq.~(\ref{e9}) originates during the calculation of the tensor $T^{\mu \nu}_{\rm QED}$ with the second Noether theorem, because the electromagnetic field $A^\mu$ is no longer treated as an external field exerting force on the Dirac field $\psi$ and $\bar{\psi}$, as it was in the calculation of the energy-momentum tensor $T^{\mu \nu}_{\rm N}$ in Eq.~(\ref{e6}).}.

The tensor $T^{\mu \nu}$ in Eq.~(\ref{e4}) constitutes a part of the energy-momentum tensor $T^{\mu \nu}_{\rm QED}$ in Eq.~(\ref{e9}). Particularly,
\begin{equation}
\label{e10}
 T^{\mu \nu}_{\rm QED} = T^{\mu \nu} +  F^{\mu}_{\phantom{\mu} \lambda} F^{\lambda \nu} - \eta^{\mu \nu}L_{\rm QED, flat}.
\end{equation}
Therefore, it is necessary to extend the Lagrangian of the GQM theory $L$ in Eq.~(\ref{e1}) with the Lagrangian of the free electromagnetic field on weakly curved space-time $L_{\rm EM}$ in Eq.~(\ref{e141}) to allow the energy-momentum tensor $T^{\mu \nu}_{\rm QED}$ to emerge in the interaction Lagrangian density of the theory $L_{\rm IG}$ instead of the tensor $T^{\mu \nu}$.

The contra-variant metric tensor $g^{\mu \nu}$ has approximately the form
\begin{equation}
\label{e12}
 g^{\mu \nu} \approx \eta^{\mu \nu} - \eta^{\mu \rho} \eta^{\nu \lambda} h_{\rho \lambda} \approx \eta^{\mu \nu} - h^{\mu \nu}.
\end{equation}
In this approximation, the second and higher order terms in the gravitational field tensor  components $h_{\mu \nu}$ are neglected. Eq.~(\ref{e12}) defines the contra-variant metric disturbance tensor $h^{\mu \nu}$.

If we substitute the metric tensor $g^{\mu \nu}$ from Eq.~(\ref{e12}) into the Lagrangian for the free electromagnetic field on curved space-time $L_{\rm EM}$ in Eq.~(\ref{e141}) and use the anti-symmetric property of the electromagnetic field tensor $F_{\mu \nu} = -F_{\nu \mu}$ and the symmetric property of the gravitational field tensor $h_{\mu \nu} = h_{\nu \mu}$, we obtain the Lagrangian of the electromagnetic field $L_{\rm EM}$ up to the first order in $h_{\mu \nu}$
\begin{equation}
\label{e15}
 L_{\rm EM} \approx -\dfrac{1}{4} \eta^{\mu \rho} \eta^{\nu \lambda} F_{\rho \lambda} F_{\mu \nu} - \dfrac{1}{2} F^{\mu \lambda} F_{\lambda}^{\phantom{\lambda} \nu} h_{\mu \nu}.
\end{equation}
The first term is equal to the Lagrangian of the electromagnetic field on flat space-time, due to Eq.~(\ref{e13}). The second term in Eq.~(\ref{e15}) describes the interaction of the electromagnetic field $A^\mu$ with the gravitational tensor field $h_{\mu \nu}$.

Since the Lagrangian $L_{\rm EM}$ is added to the Lagrangian of the GQM theory $L$ (see Eq.~(\ref{e1})), we transfer the last term on the right-hand side of Eq.~(\ref{e15}) into the interaction Lagrangian $L_{\rm IG}$, see Eq.~(\ref{e3}). There, it reconstructs part of the energy-momentum tensor $T^{\mu \nu}_{\rm QED}$ in Eq.~(\ref{e10}) related to the free electromagnetic field. As a result, we can write the Lagrangian of quantum electrodynamics on a weakly curved space-time $L$ as the sum of the Lagrangian of the same theory on flat space-time and an interaction term related to the tensor gravitational field $h_{\mu \nu}$
\begin{equation}
\label{e16}
 L = L_{\rm QED, flat} - \dfrac{1}{2}h_{\mu \nu}(T^{\mu \nu} + F^{\mu \lambda} F_{\lambda}^{\phantom{\lambda} \nu}).
\end{equation}
The tensor, which multiplies the gravitational field tensor $h_{\mu \nu}$ in the second term of Eq.~(\ref{e16}), is symmetric,  but it does not match energy-momentum tensor $T^{\mu \nu}_{\rm QED}$ of the corresponding flat space-time theory (quantum electrodynamics). In particular, the diagonal term $-\eta^{\mu \nu} L_{\rm QED, flat}$ is missing in order to fully reconstruct the energy-momentum tensor of quantum electrodynamics $T^{\mu \nu}_{\rm QED}$ in Eq.~(\ref{e10}).

In order to rectify this situation, it is necessary to work with the Lagrangian density
\begin{equation}
\label{e17}
\mathcal{L} = L \sqrt{-g},
\end{equation}
which appears under the four-dimensional volume integral for the action $S$ in Eq.~(\ref{e32}).  The Lagrangian density $\mathcal{L}$ can be considered as defining a flat space-time theory on the coordinates $x^\mu$, which are related to physical distances by the metric tensor $g_{\mu \nu}$.

\section{Interpretation of the GQM theory as a flat spacetime theory}
It has been noted~\cite{Adler2012}, that the GQM theory possesses a purely geometric term in its total Lagrangian density $\mathcal{L}$ in Eq.~(\ref{e17}) when developed from the principles of general covariance. Thus, if the weak gravitational field $h_{\mu \nu}$ was treated as a tensor field propagating on a flat space-time coupled to other fields, some terms in the resulting Euler-Lagrange equation for the Dirac field $\psi$ would be missing.

On the other hand, Feynman, Weinberg and Schwinger have pointed out, that the geometric nature of the gravitational field does not have to be taken into account~\cite{Adler2012,Weinberg1972,Feynman2002} and can be treated as an ordinary field on flat space-time as far as the gravitational field is weak. Here, we show the equivalence between the GQM theory extended with a free electromagnetic term (see Sec.~3) and an idea proposed by Feynman, Weinberg and Schwinger. At the same time we show, that the symmetric energy-momentum tensor of quantum electrodynamics $T^{\mu \nu}_{\rm QED}$ emerges in the interaction Lagrangian density $\mathcal{L}_{\rm IG}$.

We start with the calculation of the term $\sqrt{-g}$. Up to the first order in the components of the gravitational field tensor $h_{\mu \nu}$. It is equal to~\cite{Adler2012}
\begin{equation}
\label{e18}
 \sqrt{-g} \approx 1 + \dfrac{1}{2}h^\mu_{\phantom{\mu} \mu} \approx 1 + \dfrac{1}{2}\eta^{\mu \nu} h_{\mu \nu}
\end{equation}
The Lagrangian density $\mathcal{L}$ in Eq.~(\ref{e17}) can be written up to the first order in terms of gravitational field tensor $h_{\mu \nu}$ as
\begin{align}
\label{e19}
 \mathcal{L} \approx L_{\rm QED, flat} \sqrt{-g} + L_{\rm IG}.
\end{align}
Here, we consider that the theory includes the Lagrangian of the free electromagnetic field on weakly curved space-time $L_{\rm EM}$ (see Eq.~(\ref{e15})). Therefore, we can rewrite Eq.~(\ref{e19}) using Eqs.~(\ref{e18}) and (\ref{e16}) up to the first order in terms of the gravitational field tensor $h_{\mu \nu}$ as
\begin{align}
\label{e20}
 \mathcal{L} &\approx L_{\rm QED, flat}  -  \dfrac{1}{2}h_{\mu \nu}(T^{\mu \nu} + F^{\mu \lambda} F_{\lambda}^{\phantom{\lambda} \nu} - \eta^{\mu \nu} L_{\rm QED, flat}) \\
 \label{e21}
 &\approx L_{\rm QED, flat} - \dfrac{1}{2} h_{\mu \nu} T^{\mu \nu}_{\rm QED}.
\end{align}
In Eq.~(\ref{e20}), we have identified that the tensor in the product with the gravitational field tensor $h_{\mu \nu}$ is the energy-momentum tensor of quantum electrodynamics on flat space-time $T^{\mu \nu}_{\rm QED}$. Therefore, the gravitational field $h_{\mu \nu}$ has as its source $T^{\mu \nu}_{\rm QED}$ analougous to the electromagnetic  conserved four-current $J^\mu$ in flat space-time.

The Lagrangian density $\mathcal{L}$ in Eq.~(\ref{e21}) can be interpreted as describing a theory on a flat space-time $x^\mu$  involving the tensor gravitational field $h_{\mu \nu}$, the electromagnetic field $A^\mu$ and the Dirac fields $\psi$ and $\bar{\psi}$. We remark that the theory given by the Lagrangian density in Eq.~(\ref{e20}) provides the same equations of motion for the Dirac fields $\psi$ and $\bar{\psi}$ as the GQM theory. The benefit of extension of the GQM theory lies in its compatibility with the second Noether theorem. Namely, the energy-momentum tensor of the extended theory in the interaction Lagrangian density can be derived with the second Noether theorem, is symmetric, and is related to the canonical energy-momentum tensor with the Belinfante tensor. In addition, we have shown that the GQM theory proposed by Adler~et.~al. extended with the dynamics of the free electromagnetic field is equivalent to a flat space-time theory, contrary to the non-extended theory~\cite{Adler2012}.
\section{Conclusion}
The symmetric energy-momentum tensor of the GQM theory obtained by the generalization of the energy-momentum tensor of the free Dirac field theory cannot be related to the canonical energy-momentum tensor via divergence of the Belinfante tensor. Thus, they do not predict the same values of the components of the four-momentum vector. In consequence, the symmetric tensor derived by the generalization procedure is not a valid energy-momentum tensor of the GQM theory. On the other hand, the energy-momentum tensor derived by the second Noether theorem is related to the canonical energy-momentum tensor via the Belinfante tensor, but is asymmetric. The asymmetry originates from the lack of a term related to the free electromagnetic field in the Lagrangian. Particularly, we have shown that the energy-momentum tensor derived with the second Noether theorem predicts the same values for energy as the canonical energy-momentum tensor. After the extension of the Lagrangian of the GQM theory with the free electromagnetic field term on a weakly curved space-time, a symmetric energy-momentum tensor of flat space-time quantum electrodynamics appears in the interaction Lagrangian density. The tensor can be as well derived from the corresponding flat space-time theory by the second Noether theorem. It has been shown that the Lagrangian of the GQM theory~\cite{Adler2012} extended by the free electromagnetic field term is equivalent to a flat space-time theory, where the gravitational field is treated as an ordinary tensor field, in analogy to the electromagnetic field.

\section*{Declarations}
Funding to assist with the preparation of this manuscript was received from the V\v{S}B~-~Technical University of Ostrava, Faculty of Materials Science and Technology. Additional support was provided by the projects acknowledged in the Acknowledgments section.

\section*{Acknowledgements}
Support from the project No. CZ.02.01.01/00/22\_008/0004631 -- "Materials and Technologies for Sustainable Development," funded by the European Union and the state budget of the Czech Republic within the framework of the Jan Amos Komensky Operational Program and of the European Union under the REFRESH -- Research Excellence For Region Sustainability and High-tech Industries project number CZ.10.03.01/00/22\_003/0000048 via the Operational Programme Just Transition
is acknowledged.

\bibliographystyle{mrl}
\bibliography{javurek.bib}

\address{Faculty of Materials Science and Technology, V\v{S}B - Technical University of Ostrava, 17. listopadu 2172/15, 708 00 Ostrava-Poruba, Czech Republic}

\end{document}